\documentclass[final,5p,times,twocolumn]{elsarticle}

\usepackage{amssymb}
\usepackage{textgreek}
\usepackage{enumitem}
\usepackage{hyperref}
\hypersetup{
    colorlinks=true,
    linkcolor=blue,
    filecolor=magenta,      
    urlcolor=cyan,
    }
\usepackage{orcidlink}
\usepackage{comment}

\journal{Fusion Engineering and Design}

\begin{document}

\begin{frontmatter}



\title{Upgrade of the Diagnostic Neutral Beam Injector for the RFX-mod2 experiment}


\author[inst1]{Marco Barbisan\corref{cor1}\orcidlink{0000-0002-4065-3861}}
\ead{marco.barbisan@istp.cnr.it}

\affiliation[inst1]{organization={Consiglio Nazionale delle Ricerche, Istituto per la Scienza e la Tecnologia dei Plasmi (ISTP)},
            addressline={C.so Stati Uniti, 4}, 
            city={Padova},
            postcode={35127}, 
            state={Veneto},
            country={Italy}}

\author[inst2]{Marco Boldrin\orcidlink{0000-0003-0149-6266}}

\affiliation[inst2]{organization={Consorzio RFX (CNR, ENEA, INFN, Università di Padova, Acciaierie Venete SpA)},
            addressline={C.so Stati Uniti, 4}, 
            city={Padova},
            postcode={35127}, 
            state={Veneto},
            country={Italy}}
            
\author[inst3]{Luca Cinnirella\orcidlink{0009-0009-3857-7159}}

\affiliation[inst3]{organization={Centro Ricerche Fusione (CRF), Università degli studi di Padova},
            addressline={C.so Stati Uniti, 4}, 
            city={Padova},
            postcode={35127}, 
            state={Veneto},
            country={Italy}}
            
\author[inst2]{Bruno Laterza\orcidlink{0000-0002-4767-979X}}
\author[inst2]{Alberto Maistrello\orcidlink{0000-0003-4896-7328}}
\author[inst1,inst2]{Lionello Marrelli\orcidlink{0000-0001-5370-080X}}
\author[inst2]{Federico Molon\orcidlink{0000-0001-7117-6431}}
\author[inst1]{Simone Peruzzo\orcidlink{0000-0003-3626-1707}}
\author[inst1,inst2]{Cesare Taliercio\orcidlink{0000-0001-5465-9087}}
\author[inst1,inst2]{Marco Valisa\orcidlink{0000-0003-4640-5671}}
\author[inst1,inst2]{Enrico Zampiva\orcidlink{0000-0002-7516-3731}}

\cortext[cor1]{Corresponding author}

\begin{abstract}
Diagnostic Neutral Beam Injectors (DNBI), through the combined use of Charge Exchange Recombination Spectroscopy (CHERS) and Motional Stark effect diagnostics (MSE), are a well-known tool to access important information about magnetically confined plasmas, such as radial profiles of ion temperature, ion flow, impurity content and intensity and direction of the magnetic field. For this purpose, a DNBI was installed and operated in the RFX-mod experiment, which was designed to confine plasma mainly through the Reversed Field Pinch configuration. The DNBI, designed and built by the Budker Institute of Nuclear Physics (BINP), was based on a source of positive hydrogen ions, accelerated to 50 keV and for a maximum ion current of 5 A. The beam could be modulated and the maximum overall duration was 50 ms.

With the upgrade of RFX-mod to the present RFX-mod2 machine, the DNBI is being renovated to solve several power units faults and improve the overall reliability of the system. The 50 kV power supply is being improved, as well as the power supplies in the high voltage deck and its insulation transformer. Magnetic field survival tests were performed on the toroidal-core-based DC-DC converters that should power the electronic boards in a reliable way. The control system, originally based on CAMAC technology, was redesigned to be fully replaced. This contribution reviews the technical criticalities emerged in the DNBI check-up and the new solutions adopted to make the DNBI operative and more reliable.
\end{abstract}

\begin{highlights}
\item A Diagnostic Neutral Beam Injector (DNBI), in combination with optical diagnostics, can provide crucial measurements for Reversed Field Pinch Plasmas.
\item The DNBI at the RFX-mod2 experiment is in phase of extraordinary maintenance and upgrade.
\item New circuits and devices were tested. 
\item The electrical systems and the control system are being redesigned in order to be more reliable and robust.
\end{highlights}

\begin{keyword}
Neutral beam injector \sep reversed field pinch \sep Charge eXchange Recombination Spectroscopy \sep Motional Stark Effect
\end{keyword}

\end{frontmatter}


\section{Introduction}
\label{sec:intro}
The injection of a neutral beam into a magnetic confined plasma is a well established technique to probe several plasma properties. Thanks to the charge-exchange processes between beam particles and  ionized impurity atoms, these particles can emit line radiation. This light is collected by the lines of sight of the Charge eXchange Recombination Spectroscopy diagnostic (CXRS), which can measure the plasma impurity concentration (from spectral line intensities), the impurity flow (from the line Doppler shift) and the impurity temperature (from line broadening). \cite{Carolan1987,Isler1994} The beam-plasma interactions also cause the beam particles to emit H\textsubscript{\textalpha}/D\textsubscript{\textalpha} radiation; these fast particles, being immersed in the confining magnetic field, experience an electric field $\textbf{E}=\textbf{v}\times\textbf{B}$, that in turn causes the Stark splitting of the Dopppler Shifted H\textsubscript{\textalpha}/D\textsubscript{\textalpha} lines. The Motional Stark Effects (MSE) analyzes the wavelength separation of the Stark components to retrieve the intensity of $\textbf{B}$, while the identification of the polarization direction of the Stark components allows to measure the direction of $\textbf{B}$. \cite{Wroblewski1988,Levinton1989,Stratton1999,Wolf2015}

\begin{figure}[htbp]
\centering %
\includegraphics[width=8cm]{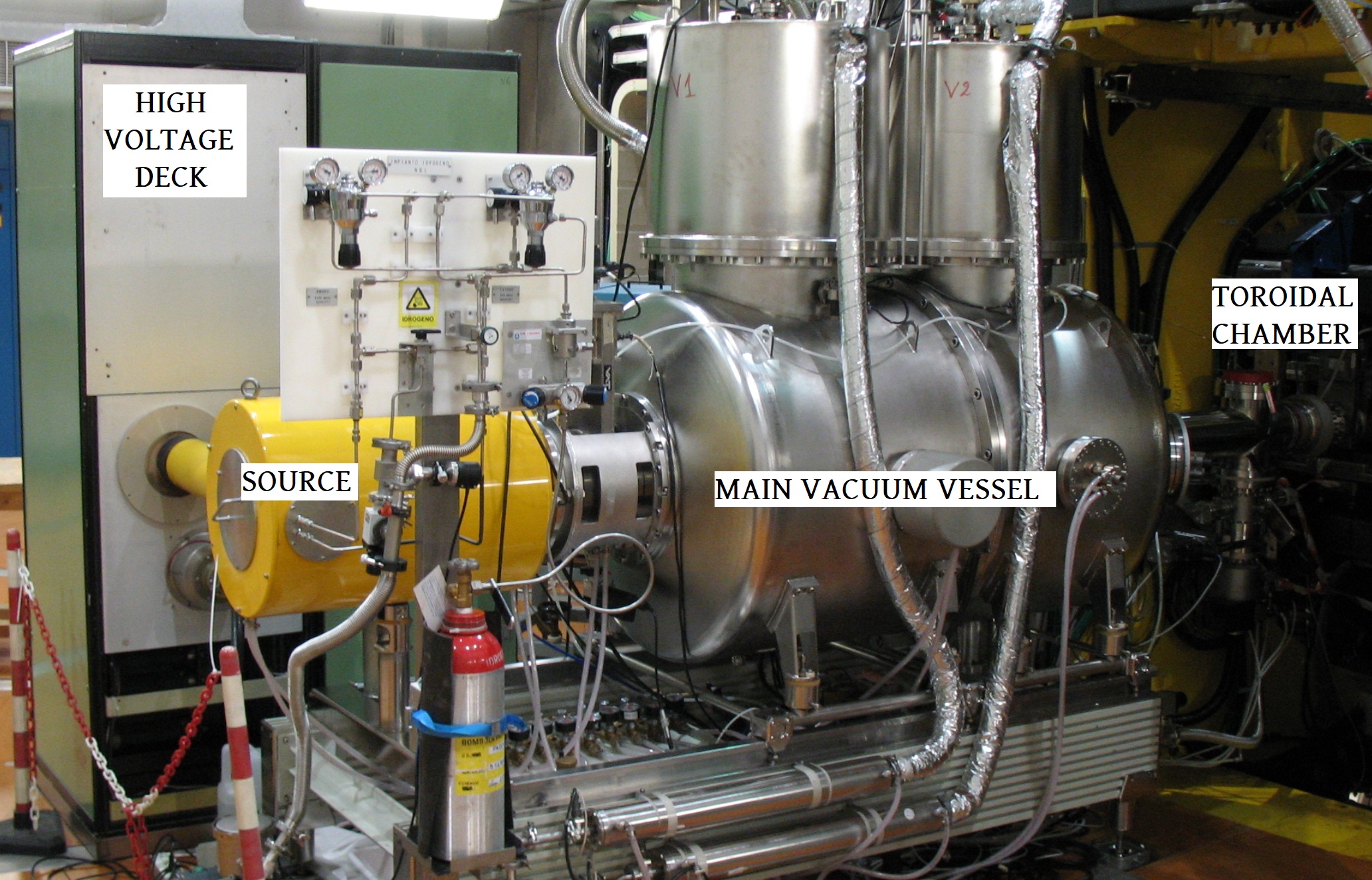}
\caption{\label{fig:DNBI} A view of the DNBI installed in RFX-mod.}
\end{figure}

Diagnostic Neutral Beam Injectors (DNBI), together with the CXRS and MSE diagnostics, have been extensively used in tokamak experiments, while less experimentation was carried out in Reversed Field Pinch (RFP) machines.\cite{Carraro1999,Kuldkepp2006,Ko2010,DenHartog2011} Being able to probe the magnetic field in the plasma core would be especially beneficial in reconstructing the complex $\textbf{B}$ RFP maps, that would otherwise rely on the only edge measurements. \cite {Bonomo2011} For this reason, the RFX-mod RFP experiment \cite{Zuin2017} in Consorzio RFX (Padua) was equipped in 2005 with a DNBI (figure \ref{fig:DNBI}), designed, built and procured by the Budker Institute for Nuclear Physics (BINP, Novosibirsk). The DNBI is based on an arc-discharge positive ion source; H\textsuperscript{+}/D\textsuperscript{+} ions are accelerated by a total potential difference of 50 kV in a four grid system, providing a maximum ion current of 5 A. Ions are then neutralized by charge exchange with a gas target. Residual ions are deflected by a magnetic field before entering the duct towards RFX-mod2. The DNBI can produce a 2 A neutral beam (measured at the DNBI exit port), of maximum duration 50 ms and maximum divergence 0.5°; the beam current could be modulated up to a maximum frequency of 250 Hz, to discriminate the beam-induced emissions from the background ones. \cite{Korepanov2004}

With the ongoing upgrade of the RFX-mod experiment to the new RFX-mod2 \cite{Marrelli2019, Peruzzo2023, Terranova2024}, the DNBI requires extraordinary maintenance and an upgrade of several power systems. The main issues are the following:
\begin{itemize}[noitemsep]
    \item The insulation transformer, powering the ion-source-related devices at 50 kV potential, is faulty.
    \item Several power and high voltage devices are aged and unsafe.
    \item The control and data acquisition system is based on an obsolete technology (CAMAC).\cite{CAMAC}
    \item The DNBI cryopumps, working in open cycle for He, are not economically sustainable.
    \item Severe beam losses were observed in the DNBI-torus chamber duct during plasma operation.
\end{itemize}
The solution of these issue is part of the NEFERTARI project (New Equipment For the Experimental Research and Technological Advancement of the RFX Infrastructure). Due to geo-political criticalities it has been impossible to collaborate with BINP, the project team has then started to fix and upgrade the DNBI in-house. Other laboratories in Europe, hosting BINP-developed DNBIs \cite{Wood2023, Listopad2024,Mysiura2024}, are affected by the same situation. The paper shares the experimental tests and the design activities on the DNBI electrical power units, that are being carried out in ISTP (Istituto per la Scienza e Tecnologia dei Plasmi) and Consorzio RFX to have an operative, safer and more robust DNBI, but also as a part of a more collective effort to keep several DNBIs in Europe into operation.

\section{The DNBI electric systems}
\label{sec:elecsystems}
The DNBI has to release hundreds of kilowatts for 50 ms every about 7 min.. A system that drains that power from the electrical net would be expensive and inefficient; instead, the energy for the most power consuming devices is slowly accumulated in banks of capacitors, which are partially discharged at the moment of the plasma and beam extraction pulses.

The DNBI power units can be divided in two main groups: those required to generate the 50 kV acceleration potential difference for the positive ions, and those required to operate the ion source. 

\begin{figure}[htbp]
\centering %
\includegraphics[width=8.5cm]{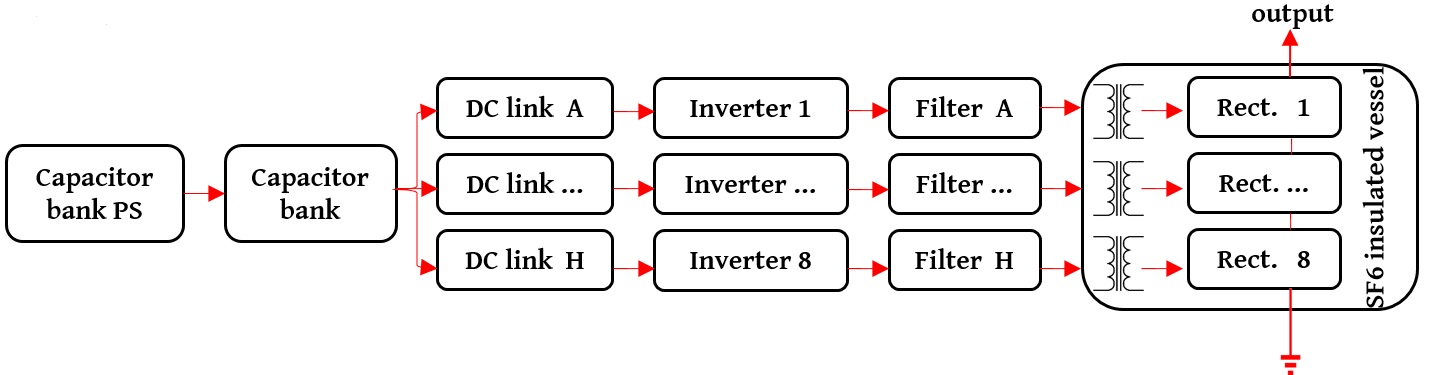}
\caption{\label{fig:50kV} Conceptual scheme of the 50 kV power supply.}
\end{figure}

Regarding the first group, the scheme of the 50 kV generation system is represented in figure \ref{fig:50kV}. The capacitor bank, 150 mF loaded at 830 V, is partially discharged through eight DC links, who regulate the input voltage for eight inverters. The filtered outputs are transferred to step-up transformers, followed by rectifiers and filters, connected in series to reach 50 kV. Transformers, rectifiers and filters are contained in a vessel, filled with SF\textsubscript{6} at the absolute pressure of 1.7 bar. The high voltage output is determined by the voltage outputs of the DC links.
\label{sec:electests}

\begin{figure}[htbp]
\centering %
\includegraphics[width=6cm]{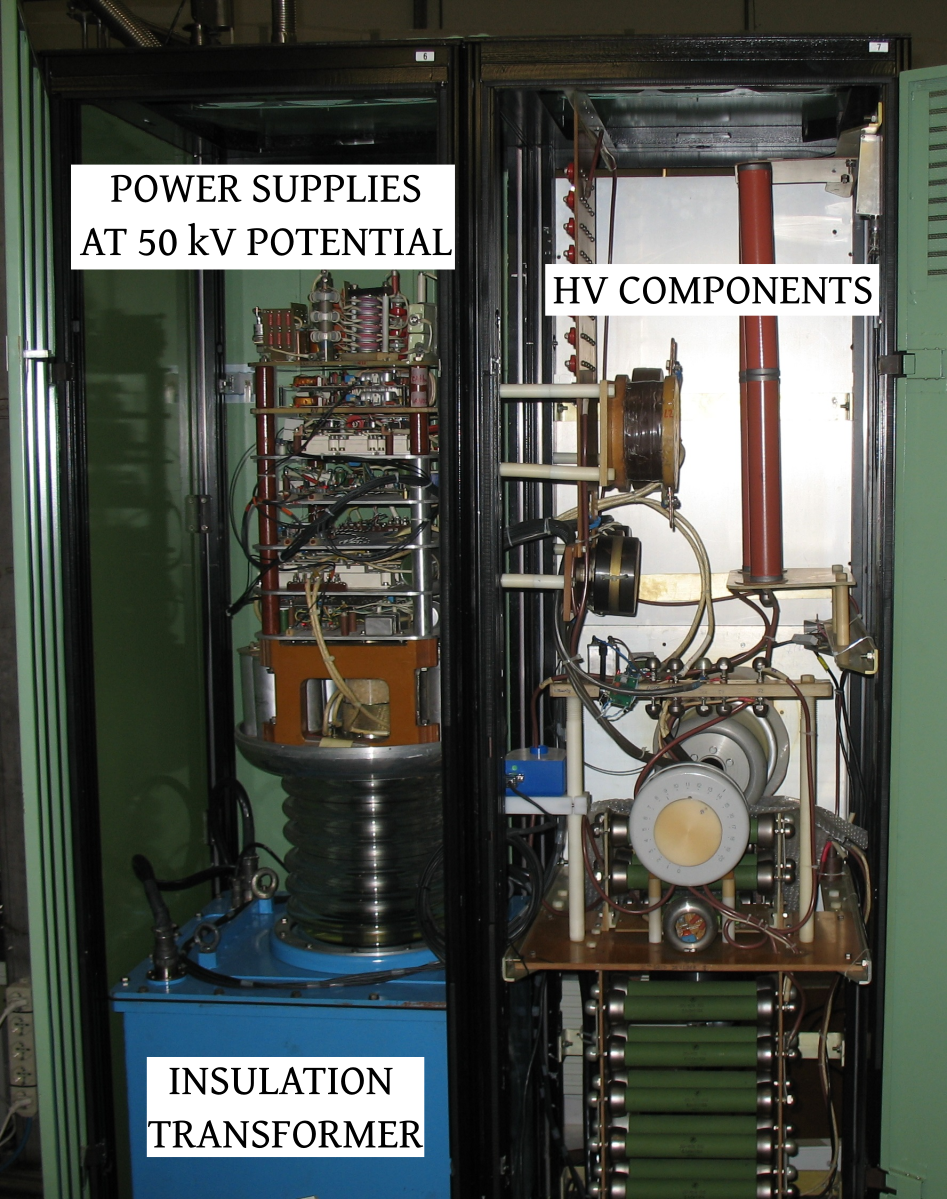}
\caption{\label{fig:HVDlayout_OLD} Present HVD layout.}
\end{figure}

Regarding the group of power units serving the ion source, it must be firstly remarked that since the final grid of the acceleration system is equipotential to the vessel, i.e. at ground potential, then the ion source and the grid facing the plasma must be held at 50 kV. As a consequence, all the source-related devices are enclosed in a High Voltage Deck (HVD) and powered through an insulation transformer. All this is installed inside a cabinet close to the ion source, together with the HV components required to derive the 2\textsuperscript{nd} grid extraction potential, to measure grids currents, to detect and protect from breakdowns and overvoltages (figure \ref{fig:HVDlayout_OLD}). 
The main power units in the HVD are:
\begin{itemize}[noitemsep]
\item The arc current power supply (ARCPS), to ignite and maintain the plasma in the ions source.
\item The magnetic insulation power supply (MIPS), whose current is used by a solenoid to generate a longitudinal magnetic field in the source. This reduces plasma losses at the ion source walls.
\item Two power supplies for the two gas valves (GVPS) that let hydrogen or deuterium enter the plasma source, from the cathode and from the anode, respectively.
\end{itemize}

Besides the two main groups of power units, other systems are also present:
\begin{itemize}[noitemsep]
\item The power supply for the suppressor grid that prevents electrons from traveling back from the beam towards the source (3rdGRIDPS).
\item The current power supply for the bending magnet that deflects non-neutralized ions before they enter the duct towards RFX-mod2 (BMPS).
\item The cooling system.
\item The pumping system (prevacuum pump, turbomolecular pump and cryopumps).
\item The beam diagnostics, as the calorimeter and four probes, placed at cardinal points at the entrance of the beam duct, to detect beam misalignments. 
\end{itemize}
Table \ref{tab:PS} gives an overview of the basic specifications for the main DNBI power units. For ARCPS and GVPS, when a double value is indicated, the first one refers to the initial instant for arc ignition or for opening the valve, while the second one refers to the rest of the pulse duration.

\begin{table}[htbp]
\footnotesize
\centering
\caption{Maximum ratings of the main DNBI power supplies.}
\label{tab:PS}
\begin{tabular}{lllll}
\ Power & Voltage& Current& Rise-fall&Modulation\\
\ supply &  (V) &(A) &time (ms)& (Hz)\\
\hline
50 kV system	&50000&8&0.2&250\\
3rdGRIDPS	&700&1&0.2&250\\
BMPS	&20&150&2000&-\\
ARCPS	&2000/150&500&0.2&250\\
MIPS	&35&15&0.2&-\\
GVPS	&250/36&0.4/10&0.2&-\\
\hline
\end{tabular}
\end{table}

\section{Electrical tests}

\subsection{High voltage tests}

\begin{figure}[htbp]
\centering %
\includegraphics[width=8.5cm]{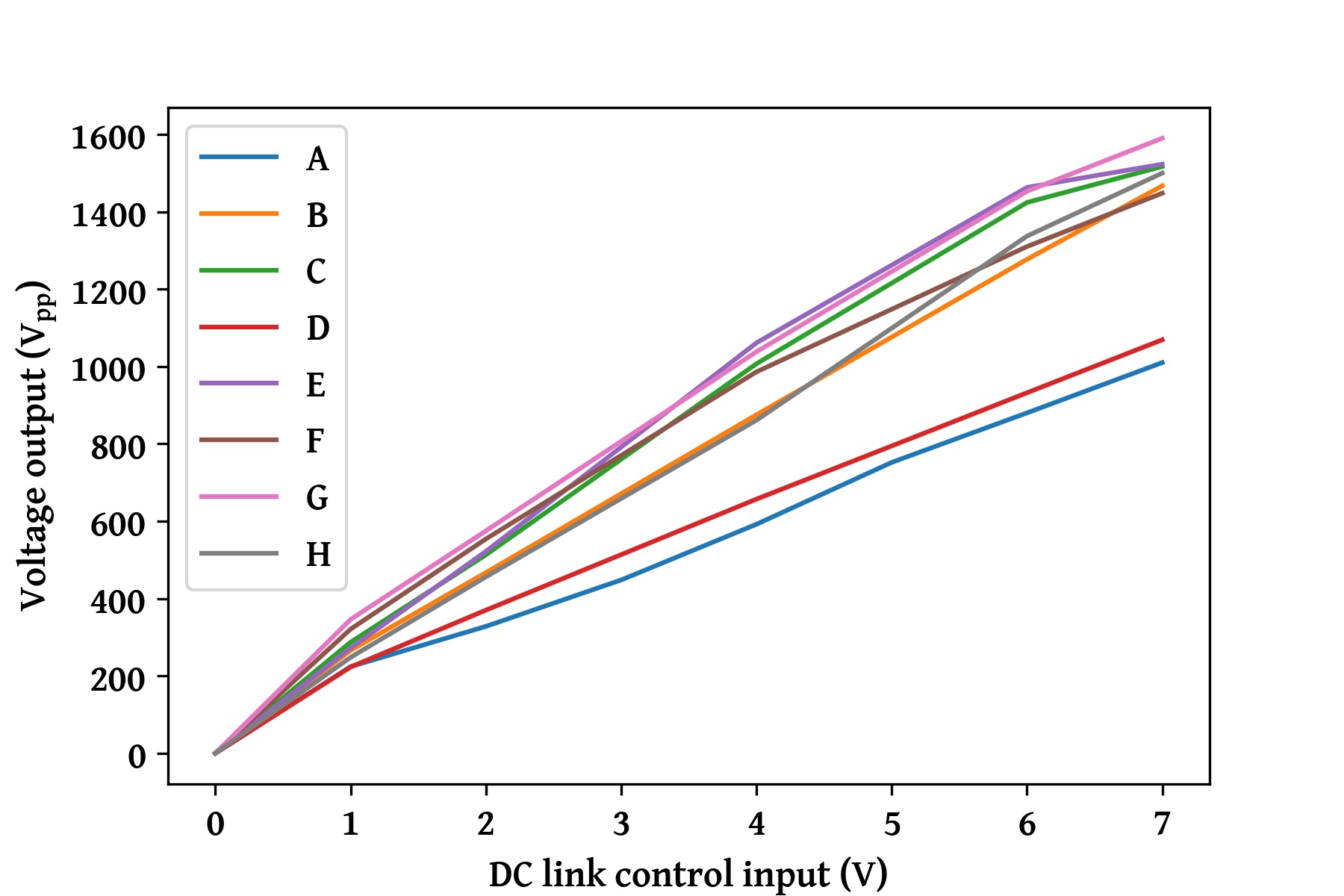}
\caption{\label{fig:inverter} Peak-to-peak voltage output of each inverter line for 50 kV generation, as a function of the analog control input. Measurements are performed on a 150 {\textOmega} resistive load.}
\end{figure}

The restoration of the chain of devices aimed at producing the 50 kV acceleration voltage involved several activities: a slow reforming of the electrolytic capacitors, the restoration of the power supply to load the capacitor bank, a careful check of all wirings between boards and racks, progressive load tests on each DC-link and inverter, and finally SF\textsubscript{6} leak tests and replacement of the gas inside the power supply vessel.
Before coupling the inverter and filters to the respective step-up transformers, their output was tested one by one on a 150 {\textOmega} resistive load (max. 1 kW transient, 50 W steady state). The peak-to-peak AC output voltage of the eight inverters is shown in figure \ref{fig:inverter} as a function of the DC link analog control input. Two sets among the eight ones gave an output voltage about 30 \% lower than the other ones; wiring losses on the analog input line and different internal settings of the control boards were excluded as possible causes of this issue. The overall aging of the electronics is considered responsible of the issue; this suggests to at least replace the low voltage control electronics, and make its design more robust. The DNBI cabinets are placed in same hall of the RFX-mod2 toroidal chamber: frequent repairs would not be tolerated during experimental sessions.

In the unavailability of a resistor to mimic the acceleration system load (50 kV, about 5 A), the high voltage output of the entire system was tested on the resistive divider (bottom right corner of figure \ref{fig:HVDlayout_OLD}) which is used to derive the potential of the second grid. The total resistance of the divider amounts to 55 k{\textOmega}. In these conditions,  the entire system was successfully able to reach the output voltage of 50 kV for at least 50 ms, even with the low performance of inverters A and D. 0 \%-100 \% output modulation tests were performed, showing a rise/fall time below 1 ms, which is adequate for the DNBI operation.

\begin{figure*}[btph]
\centering %
\includegraphics[width=18cm]{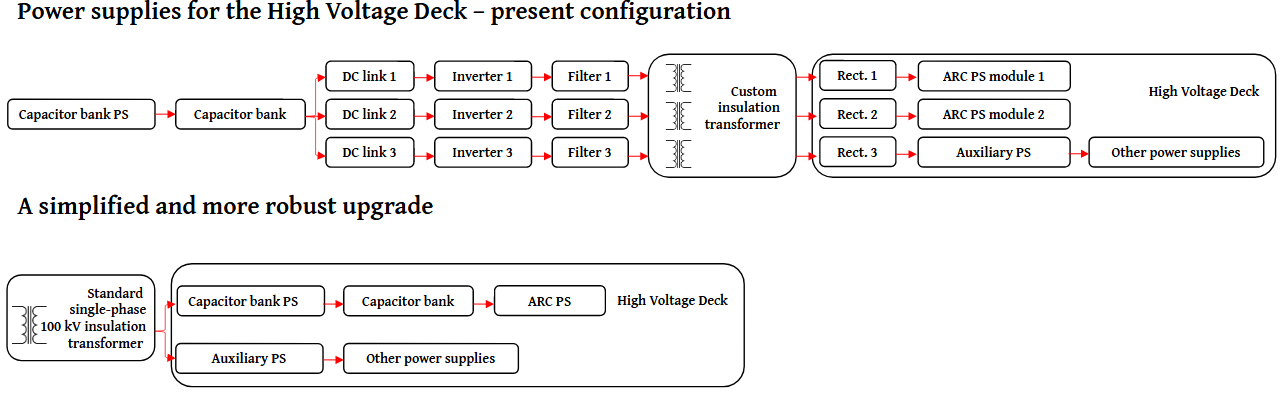}
\caption{\label{fig:HVDimprov} Conceptual schemes of the present and future HVDs.}
\end{figure*}

\subsection{IGBT control}
Insulated-Gate Bipolar Transistors (IGBTs) are the core of DC links and inverters. To manage the IGBT switching, a 160 kHz clock signal is generated in the main control board. In every group of 8 digital pulses, every pulse is sequentially sent to the controller a different DC link-Inverter line; the received pulses set the frequency (20 kHz) of the IGBT switching. This method allows to minimize the overall high voltage ripple.
The aging of electronics and the documentation of past operational issues suggested to redesign the control boards for the DC link-inverter lines. To have a more inspectable system, the core of the new control boards will be an Espressif ESP32-C3 microcontroller, which will be able to communicate with the control system in Modbus standard. \cite{ESP32-C3} Main duty of the microcontroller will be to retransmit the clock pulse to the IGBT drivers, adjusting the duty cycle as a function of an analog control waveform and of an opto-isolated analog feedback from the DC link. Both analog inputs will be acquired by a Microchip MCP3202 ADC (12 bit, 100 kS/s total).\cite{MCP3202} The duty cycle will determine the DC link output voltage; the microcontroller should be able to adjust it in between every 20 kHz pulse. Bench tests verified that the output pulse from the microcontroller has a delay between 50 ns and 100 ns with respect to the input pulse; moreover, the jitter of the output pulse is less than 50 ns. These values are considered adequate.

The microcontroller is supposed to control the IGBTs of the original DNBI (Semikron SKM200GB128D \cite{SKM200GB128D}) through a gate driver, which should provide adequate insulation between control and IGBT and manage IGBT protection systems. The original driver (Powerex M57962L \cite{M57962L}) will be substituted by the Texas Instruments ISO5451 \cite{ISO5451}, featuring a better short propagation delay (76 ns vs 1.0 {\textmu}s), a better insulation voltage (5700 V\textsubscript{RMS} vs 2500 V\textsubscript{RMS} AC) and an insulated fault pin. For a preliminar test, the driver interfaced the microcontroller and an IGBT according to the recommended scheme in figure 10-3 of ref. \cite{ISO5451}. The controlled IGBT could succesfully drive a 80 A current (foreseen in the DNBI devices) on a 0.24 {\textOmega} resistive load, and repeatedly open the circuit. A 37.5 {\textOmega} resistor was put in parallel to the collector and the emitter of the IGBT, to limit overvoltages when the IGBT opened the load circuit. The desaturation and active Miller clamp protection features of the ISO5451 were succesfully tested.

\subsection{Magnetic field survival tests}

The DNBI is equipped with several lab-built low voltage power supplies, that serve control and power devices. To improve the overall reliability and maintainability of the system, these power supplies will be substituted with commercial DC-DC converters. These devices, as well as their predecessors, could be affected by the magnetic fields around the RFX-mod2 machine: they all employ toroidal magnetic cores, that could get saturated. Significant variations in the available voltage or current may lead to unpredicatble behaviour of the main DNBI devices; moreover, identifying these issues during the short RFX-mod2 pulses (around 1 s) and with limited access to the hall may increase the unavailability of the DNBI in case of electric faults.

To prevent this risk, several commercial DC-DC converters, listed in table \ref{tab:Bfield}, were tested at 130 mT, i.e. twice the magnetic field intensity expected at the position of these devices. Doubling the magnetic field was conservatively chosen to account for possible future modifications of RFX-mod2 solenoids and related power supplies, and to somehow account for the cumulative stress DC-DC converters would undergo during their lifetime.  The converters were tested at about 2/3 of their respective maximum output load, with the magnetic field at three possible orthogonal axes. The converter 1 proved to stand 130 mT at 10 magnetic field pulses, 2 s long, or at a single 10 s magnetic field pulse (time was limited due to the heating on the coil used to produce the field). These times are relatively long compared to the DNBI operation: the plasma phase should last 100 ms, the beam phase 50 ms. It was experimentally determined that the converter 1 can reliably switch on only if powered before the beginning of a magnetic field pulse. All the other converters stopped working under magnetic field, some in reversible ways, while others were permanently damaged. Enclosing the converters 2-7 in transformer core foils improved their resistance to intermediate magnetic field intensities, but not allowed them to properly work along all three possible magnetic field directions.

A complete characterization of the DC-DC converter 1 would also require to test it under rapidly changing magnetic fields, as in a RFX-mod2 shot. The available current generator could not allow such tests. The only possibility to perform this kind of test will likely come during the commissioning of RFX-mod2 solenoids. If a spare DC-DC converter 1 will fail to operate under those conditions, magnetic shields will be placed in the DNBI electronic boards before they are put into operation.

\begin{table}[htbp]
\footnotesize
\centering
\caption{DC-DC converters tested at 130 mT magnetic field.}
\label{tab:Bfield}
\begin{tabular}{lll}
\# & Model&Ref.\\
\hline
1	&TRACO POWER TEN 5-2423&\cite{TEN5_2423}\\
\hline
2	&TRACO POWER THM 3-0523&\cite{THM_3_0523}\\
\hline
3	&TEXAS INTR. DCP022405DP&\cite{DCP022405DP}\\
\hline
4	&MURATA NMA1212SC&\cite{NMA1212SC}\\
\hline
5	&MURATA CRE1S0505S3C&\cite{CRE1S0505S3C}\\
\hline
6	&TEXAS INSTR. DCP010505DBP&\cite{DCP010505DBP}\\
\hline
7	&MURATA MGJ2D151505MPC-R7&\cite{MGJ2D151505MPC-R7}\\
\hline
\end{tabular}
\end{table}

\section{Upgrade of electrical systems}
\label{sec:upelectrical}
\subsection{High Voltage Deck power transfer}
One of the main electrical issues with the present DNBI is the faulty insulation transformer that powers the devices inside the HVD. As schematized in figure \ref{fig:HVDimprov}, there are actually three insulation transformers inside the oil container; they are served by DC link-inverter couples, which drain charge from a common capacitor bank. The outputs of two insulation transformers power the arc current power supply, necessary to ignite and sustain the plasma. The third insulation transformer powers the auxiliary power supplies, eg. the magnetic insulation power supply and the power supplies for the gas valves.

\begin{figure}[htbp]
\centering %
\includegraphics[width=8.5cm]{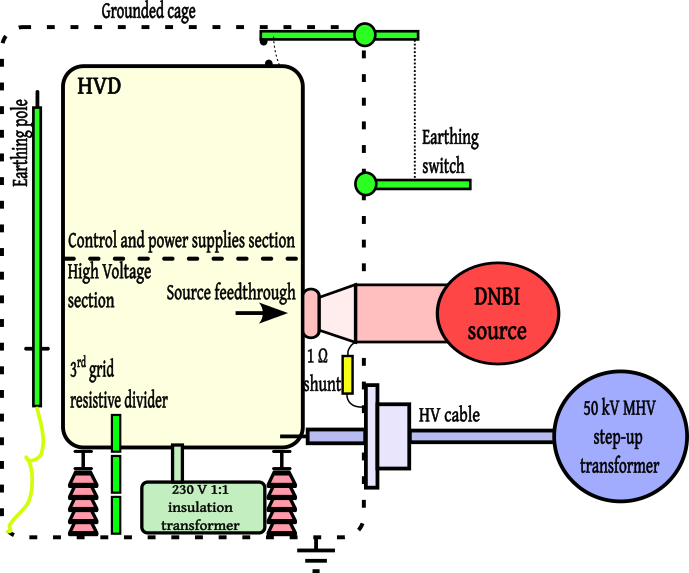}
\caption{\label{fig:HVDlayout_NEW} Conceptual design for the new HVD.}
\end{figure}
Inspectioning and repairing a custom triple insulation transformer may have uncertain results; replacing it in the present form would be unlikely and expensive (scarcity of manufacturers in the market). It was then decided to change the powering scheme of the HVD. As shown in figure \ref{fig:HVDimprov}, the new scheme foresees a simple single-phase 100 kV insulation transformer. Inside the HVD, it will directly serve the low power auxiliary power supplies, as well as a commercial 500 V DC power supply. The capacitor bank will be moved inside the HVD and kept loaded by this power supply. The capacitor bank will in turn serve the arc current power supply. As evident from figure \ref{fig:HVDimprov}, this solution will allow not only to subtitute the insulation transformer, but also to greatly simplify the electrical power systems, hopefully making them more robust and reliable.

Figure \ref{fig:HVDlayout_NEW} shows a preliminar layout of the new HVD. A commercial cabinet is kept at 50 cm height from the ground. The insulation transformer will be installed in between. The 50 kV output will be linked to the chassis of the cabinet, and used for all HV components; these will be completely rearranged and substituted, because of the number of unexpected arcs occured during HV tests. The ion source inputs will be routed through coaxial pipes. While the HV components will be hosted in the lower part of the HVD, the remaining devices (control and data acquisition, capacitor bank, arc power supply, etc.) will stay in the upper half, protected by arcs and related electromagnetic interference by a metal shield. At last, the whole HVD will be made inaccessible by a grounded cage.

\subsection{Arc current power supply}
\begin{figure}[htbp]
\centering %
\includegraphics[width=8.5cm]{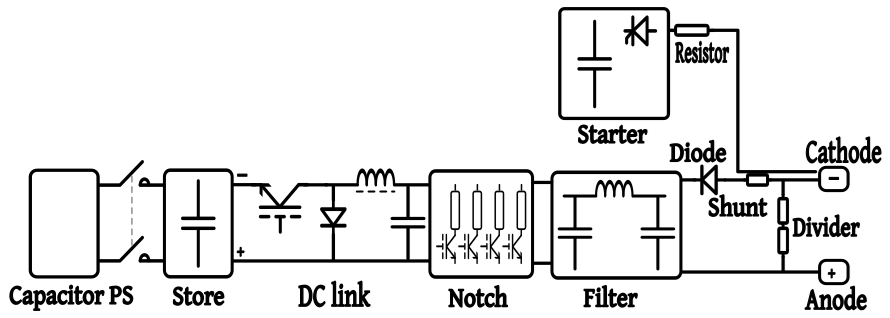}
\caption{\label{fig:ARCPS} Conceptual design of the arc power supply.}
\end{figure}

The arc power supply, schematized in figure \ref{fig:ARCPS}, is required to initiate and sustain the plasma. The maximum output ratings are 150 V, 500 A; the arc ignition is expected to require less than 2 kV. The ignition pulse will be provided by a dedicated 2 nF - 5 kV capacitor kept in charge by a power supply; the capacitor discharge will be triggered by a series of thyristors. A 2 k{\textOmega} resistor will limit the discharge current. The main arc current will instead be derived from the charge of the capacitor bank in the HVD. A DC link will tailor the output according the analog waveform given by the acquisition system, and according to the feedback on the output current provided by a shunt resistor.

In case of breakdowns in the beam acceleration system, the acceleration voltage will be totally zeroed for few milliseconds. Similarly, the arc current output will have to be notched to a minimum value, in order to not stop the arc. For a robust control of the notch, the current reduction factor will be set by four digital inputs, triggering four notch systems, each one adding a lower and lower resistance in parallel to the output. A LCR net will finally filter the arc current output. The arc voltage will be measured through a resistive divider, while a diode will prevent current inversions from entering the power supply (eg. the ignition pulse).

\section{Upgrade of the control system}
\label{sec:upcontrol}

The present control and data acquisition (CODAQ) system is based on the obsolete CAMAC standard \cite{CAMAC}, which is not reliably maintainable anymore; no spare parts would be available, and the aged instrumentation is prone to electrical faults. The CAMAC system is hosted in a single crate in the cabinets at ground potential, that also host the capacitor bank power supply, the capacitor banks themselves, DC link and inverters, etc.. The few inputs/outputs from and to the HVD were dispatched by means of optical fibers, one per channel. The CAMAC system, in terms of active channels, includes 57 digital inputs, 25 digital outputs, 17 analog inputs and 9 analog outputs at few tens of samples per second. Moreover, 12 fast analog inputs are acquired up to 2 MS/s, and 16 timing signals (digital outputs) are available at 1 {\textmu}s time resolution. Analog conversion has 16 bit resolution, with the exception of fast digital inputs (12 bits).

\begin{figure}[htbp]
\centering %
\includegraphics[width=8.5cm]{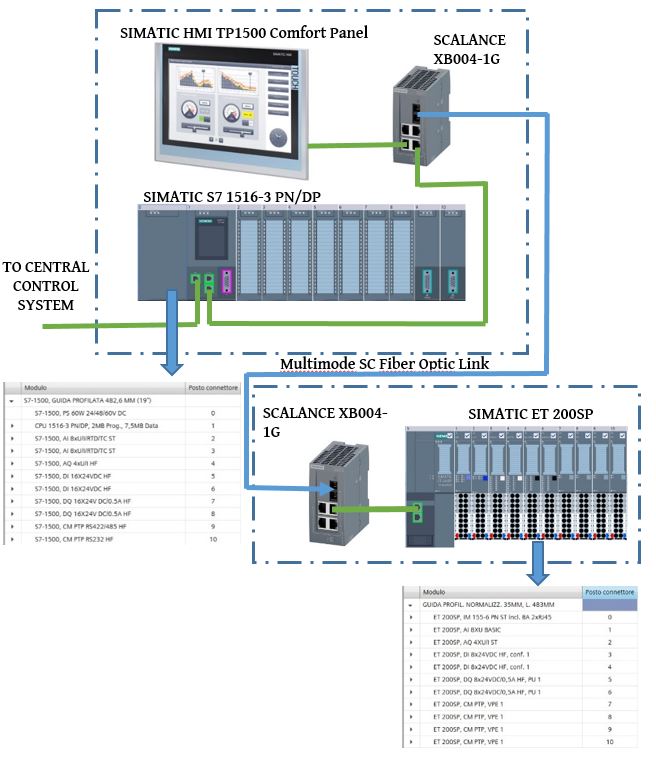}
\caption{\label{fig:PLC} PLC control for the upgraded DNBI. \cite{PLCMAIN,PLCperiph,touch,fiberconn}}
\end{figure}

While essentially maintaining the CAMAC performaces in terms of time resolution and analog-digital conversion resolution, the design of the new CODAC was based on modern devices for the standard industrial automation, for better reliability and maintainability. More specifically, the new CODAC will be composed by two subsystems. The first one will be based on a PLC, for slow analog and digital inputs/outputs (up to 50 S/s, 0-24V, 16 bit resolution) and especially for all signals involved in machine or human safety. The PLC will also manage the pumping system, which has been separately controlled up to now by a {\textmu}PLC SIEMENS “LOGO! 12/24RC” \cite{LOGO} alongside the CAMAC devices. The devices composing the new PLC system are schematized in figure \ref{fig:PLC}. The PLC will be hosted in the cabinets at ground potential, together with a touch panel for the inspection of system data or for the manual operation of the DNBI during preliminary tests or maintenance operations. For the chosen devices, an insulation voltage of 707 V DC is guaranteed; the inputs/outputs related to the 830 V capacitor banks, DC links and inverters, will be further protected by optoisolators. The PLC will be connected via optical fiber to a peripheral in the HVD, for the inputs/outputs of the devices at 50 kV potential. This solution will allow to handle several digital inputs and outputs with a single fiber, but most of all, it will allow to reliably transfer analog signals between HVD and ground cabinets, without facing the risks of transducing voltage into light intensity and vice versa in an analog way.

The second subsystem will be instead devoted to fast data acquisition (DAQ) and for fast analog and digital outputs. Following a similar scheme with respect to the PLC, a D-tAcq ACQ2106 modular data acquisition system will be used for the signals in the ground cabinets, while a D-tAcq ACQ1002R will be installed in the HVD, with data transfer via fiber optic link. \cite{DTACQ_GND,DTACQ_HVD} These devices will allow an easy integration of data storage in the MDSplus software, which was adopted in the whole RFX-mod machine as well as in RFX-mod2.\cite{martini:icalepcs2023-tupdp042} DAQ analog inputs and outputs will have 16 bit resolution and ±10 V range; both analog and digital inputs/outputs will be sampled at 1 MS/s. Similarly to the PLC, having a peripheral in the HVD will allow to reliably transfer fast analog signals across a 50 kV potential difference.

At last, an overview of the number of signals to be handled in the DNBI is provided in table \ref{tab:IOs}. PLC/GND and DAQ/GND refer to the PLC and fast acquisition for the cabinets at ground potential and for the other systems at ground potential (eg. the pumping system, the diagnostic calorimeter, etc.).

\begin{table}[htbp]
\footnotesize
\centering
\caption{Input and outputs for the control and data acquisition system.}
\label{tab:IOs}
\begin{tabular}{lcccc}
 & DAQ GND & DAQ HVD & PLC GND & PLC HVD \\
\hline
RS232, Modbus, etc.&1&0&10&6\\
\hline
Analog Input, slow&16&0&9&2\\
\hline
Analog input, fast&7	&9&	0&	0\\
\hline
Analog output, slow&0	&0	&3	&2\\
\hline
Analog output, fast&2	&2&	0	&0\\
\hline
Digital Input, slow&0	&0	&55	&11\\
\hline
Digital input, fast&3	&0	&1	&0\\
\hline
Digital output, slow&0	&0	&22	&14\\
\hline
Digital output, fast&3	&3	&0	&0\\
\hline
\end{tabular}
\end{table}

\section{Conclusions}
\label{sec:conclusions}
Diagnostic Neutral Beam Injectors are fundamental to retrieve several properties of the plasma core by means of spectroscopy. Given the impossibility of refurbishing the RFX-mod2 DNBI from its manufacturer to solve several electrical issues, several tests showed that the system power devices that are responsible for the acceleration system can still generate 50 kV. This was an important milestone in determining the overall feasibility and convenience of the DNBI refurbishment.

The control of IGBTs, used in the inverters for 50 kV generation and in several other power units, will feature microcontrollers for the feedback adjustment of the pulse width modulation. This will give better inspectability and reliability than the old analog feedback circuitry. The new IGBT drivers will provide a shorter propagation delay and a higher insulation voltage. 

Magnetic field survival tests were performed on candidate DC-DC converters that should supply low voltage for the electronic boards. While DC-DC converters are the most frequent kind of device that could be affected by magnetic fields, other inductor nuclei may be found in commercial devices that cannot be easily inspected or modified, eg. PLC devices, cryopump control boards, etc.; some of them were not available for the tests and would be too expensive to be damaged in a test. In these cases, the devices will be kept far from too intense magnetic fields wherever possible, otherwise magnetic shields will be used. 

The new design of the HVD will ease the integration of power units in standard racks, for better maintainability. The new HVD powering system will significantly reduce the number of power units and the need for a custom oil insulation transformer. 

The new control system will be maintainable and better interfaceable with modern hardware and software for automation industry. The PLC will have full control over the pumping system, differently from the CAMAC boards, that had to be integrated by a separate control system. The choice of using PLC and fast acquisition peripherals inside the HVD will allow for more freedom in the number of inputs/output, with the use of just two fibers. It will be possible to reliably tranfer analog signals to and from the HVD. 

In parallel to the upgrades to the electrical power units, the DNBI cryopumps are planned to be replaced with new closed-cycle devices, for better maintainability and economical sustainability (affected by liquid He costs). Moreover, the duct between the DNBI and RFX-mod2 is being redesigned to minimize the gas pressure; the beam reionization caused by the gas target, in combination with magnetic fields of several tens of milliTeslas, would indeed lead to losing most of beam current in the duct. The gas flux simulations and the new design of the pumping system and of the beam duct will be discussed in a future paper. As final remark, the authors wish that future geo-political conditions will re-establish peace and international freedom in research communications.  

\section*{Data availability}
The data and Python code to generate figure \ref{fig:inverter} and the source code to test the ESP32-C3 microcontrollers are available at ref. \cite{dataset}.

\section*{Acknowledgements}
This work has been funded by the European Union - NextGenerationEU (Mission 4, Component 2, CUP B53C22003070006). Views and opinions expressed are however those of the author(s) only and do not necessarily reflect those of the European Union or the European Commission. Neither the European Union nor the European Commission can be held responsible for them.

The authors thank L. Baseggio, D. Fasolo, M. Fincato, L. Franchin, G. Passalacqua and E. Zerbetto for their support in the activities of the project.

 \bibliographystyle{JHEP}
 \bibliography{cas-refs}





\end{document}